\documentclass[11pt]{article}
\usepackage[colorlinks=true]{hyperref}
\hypersetup{colorlinks=true,linkcolor=teal,citecolor=blue,urlcolor=blue}
\usepackage{graphicx}
\usepackage{amsmath}
\usepackage{amssymb}
\usepackage{latexsym}
\usepackage{color}
\usepackage{float}
\usepackage{subfigure}
\usepackage{pst-all}
\usepackage{pstricks,pst-node,pst-text,pst-3d}
\usepackage{ifpdf}
\usepackage{multimedia}
\usepackage{lmodern}
\usepackage{cite}
\usepackage{dsfont}
\usepackage{bbm}
\usepackage{lipsum}
\usepackage{url}
\usepackage{textcomp}
\usepackage{bbold}
\usepackage{epstopdf}
\usepackage{epsfig}
\usepackage{mathtools}
\usepackage{nccmath}
 \usepackage[normalem]{ulem}
\usepackage{ifpdf}
\ifpdf
\else
  \usepackage{pstricks}
\fi
  \hypersetup{pdfstartview=XYZ}
\newcommand{\bra}{\begin{array}}
\newcommand{\era}{\end{array}}
\newcommand{\beq}{\begin{equation}}
\newcommand{\eeq}{\end{equation}}
\newcommand{\beqar}{\begin{eqnarray}}
\newcommand{\eeqar}{\end{eqnarray}}

\def\BC{\bb C}
\def\_\BC{\bbi C}

\def\( {\left(}
\def\) {\right)}
\def\[ {\left[}
\def\] {\right]}
\def\no2 {{\textstyle{n\over 2}}}

\def\dag {{\dagger}}

\begin{document}
\thispagestyle{empty}
\begin{center}

\vspace{1.8cm}
 \renewcommand{\thefootnote}{\fnsymbol{footnote}}


 {\Large {\bf The effects of detuning on entropic uncertainty bound and quantum correlations in dissipative environment  }}\\

\vspace{1.5cm} {\bf Shahram Mehrmanesh}$^{1}$, {\bf Maryam Hadipour}$^{1}$,  {\bf Soroush Haseli}$^{1}${\footnote { email: {\sf
soroush.haseli@uut.ac.ir}}}\\
\vspace{0.5cm}

$^{1}$ {\it  Faculty of Physics, Urmia University of Technology, Urmia, Iran}\\ [0.3em]

\end{center}
\baselineskip=18pt
\medskip
\vspace{3cm}
\begin{abstract}
One of the fundamental arguments in quantum information theory is the uncertainty principle. In accordance with this principle, two incompatible observables cannot be measured with high precision at the same time. In this work, we will use the entropic uncertainty relation in the presence of quantum memory.  Considering a dissipative environment, the effects of  the detuning between the transition frequency of a quantum memory and the center frequency of a cavity on entrpic uncertainty bound and quantum correlation between quantum memory and measured particle will be studied. It is shown that by increasing the detuning, quantum correlation is maintained. As a result, due to the inverse relationship between the uncertainty bound and quantum correlation, the measurement results is guessed more accurately.
\end{abstract}
\noindent {\it Keywords:} Entropic uncertainty bound; Detuning; Quantum correlations.
\vspace{1cm}

\newpage
 \renewcommand{\thefootnote}{*}
\section{Introduction}
Classically, measurement errors are caused by inaccuracies in measuring devices. Quantum theory prohibits measuring two incompatible observables at the same time, regardless of the accuracy of the instrument. Quantum theory explains it by  uncertainty principle. In $1927$, Heisenberg proposed the first uncertainty principle for the two incompatible observables : position and momentum \cite{heyzenberg}.  According to Kennard \cite{Kenard}, Heisenberg's uncertainty principle can be formalized as $\Delta \hat{x} \Delta \hat{p}_x \geq \hbar /2$, where $\Delta \hat{x}$ and $\Delta \hat{p}_x$ are  standard deviations
of the position and momentum respectively. Heisenberg's uncertainty relation was generalized by Robertson and Schrodinger for any two incompatible observables $\hat{Q}$ and $\hat{R}$. As a result, one has the following uncertainty relation for arbitrary quantum state $\vert \psi \rangle$
\begin{equation}\label{robertson}
\Delta \hat{Q} \Delta \hat{R} \geq \frac{1}{2} \vert \langle \psi \vert [\hat{Q},\hat{R}] \vert \psi \rangle \vert,
\end{equation} 
where $\Delta \hat{\square}= \sqrt{\langle \psi \vert \hat{\square}^{2} \vert \psi \rangle -\langle \psi \vert \hat{\square} \vert \psi \rangle^{2} }$ is the standard deviation of the observable $\hat{\square}$ and $[\hat{Q}, \hat{R}]=\hat{Q}\hat{R}-\hat{R}\hat{Q}$. It is also possible to determine the uncertainty relation for any two general observables using entropic measures \cite{Wehner}. At first, Kraus \cite{kraus} introduced the entropic uncertainty relation, which was later proved by Maassen and Uffink \cite{masen}. As shown below, it takes the form of
\begin{equation}
H(\hat{Q})+H(\hat{R}) \geq \log_2 \frac{1}{c},
\end{equation}
where $H(\hat{X})=- \sum_x p_x \log_2 p_x$ is the Shanon entropy of the measured observable $\hat{X} \in \lbrace \hat{Q}, \hat{R}\rbrace$ and $p_x$ is the probability hat the observable measurement outcome  is equal to $x$. Complementarity between the observables Q,R is quantified by the coefficient $c=\max_{i,j} \vert \langle q_i \vert r_j \rangle \vert^{2}$, where $\vert q_i \rangle$ and $\vert r_j \rangle$ are eigenvectors of $\hat{Q}$ and $\hat{R}$, respectively. In Ref.\cite{berta}, the tighter bound has been introduced for mixed state $\rho$ as 
\begin{equation}
H(\hat{Q})+H(\hat{R}) \geq \log_2 \frac{1}{c}+S(\hat{\rho}),
\end{equation}   
where $S(\rho)=-tr(\rho \log_2 \rho)$ is the Von Neumann entropy of the the state with density operator $\rho$. It has been shown by Berta et al. that  entropic uncertainty relation can be  expressed more comprehensively using an additional particle that is correlated with the measured particle \cite{berta2}. The additional particle is known as quantum memory. Their results indicate that Bob's uncertainty about Alice's measurement outcome follows the following uncertainty relation
\begin{equation}
S(\hat{Q} \vert B)+S(\hat{R} \vert B) \geq \log_2 \frac{1}{c} + S(A \vert B),
\end{equation}
where $S(A \vert B)=S(AB)-S(B)$ is the conditional von Neumann entropy, $S(X \vert B)$ is the post measurement conditional von Neumann entropy when Alice measures $X \in \lbrace \hat{Q}, \hat{R}\rbrace$. Note that post measurement state in $X$ basis $\lbrace \vert x_i \rangle\rbrace$ can be written as
\begin{equation}
\rho^{XB}=\sum_i \Pi_x^A \rho^{AB}\Pi_x^A ,
\end{equation} 
where $\Pi_x^A=\vert x_i \rangle \langle x_i \vert \otimes \mathrm{I}$ and $\mathrm{I}$ is the identity operator. Numerous studies have been conducted about the entropic uncertainty relation with quantum memory EUR-QM \cite{8,a,b,c,d,e,f,g,h,i,j,k,9,10,11,12,13,14,15,16,17,18,19,20,21,22,23,24,25,26,27,28,29,30,31,32,33,34,35,36,37,38,39,40,41,42,43,44,45,46,47,48,49,50,51,52,53,54,55,56,57,58,59,60,61,62,63,64,65}
. So far, Adabi et al. have been proposed the tightest bound as follow \cite{61}
\begin{equation}\label{Adabi}
S(Q \mid B)+S(R \mid B) \geq \log _2 \frac{1}{c}+S(A \mid B)+\max \{0, \delta\}
\end{equation}
with $\delta=I(A ; B)-(I(Q ; B)+I(R ; B))$. $I(X;B)=S(B)-\sum_i p_{x_i} S(\rho_{x_i}^{B})$ is the Hollevo quantity with $X \in \lbrace \hat{Q}, \hat{R}\rbrace$, where $p_{x_i}=\operatorname{tr}_{A B}\left(\Pi_x^A \rho^{A B} \Pi_x^A\right)$ is the probability of $x$-th outcome and $\rho_x^B=\operatorname{tr}_A\left(\Pi_x^A \rho^{A B} \Pi_x^A\right) / p_x$ is the post measurement state of the Bob after measuring $X$. We refer to the right-hand side of inequality  (\ref{Adabi})   as entropic uncertainty bound from now on.

In recent years, quantum correlations have been extensively studied due to their fundamental role in quantum information theory \cite{66,67,68,69,70,71,72,73}. Quantum correlations allow for a variety of applications in quantum information theory, including quantum teleportation \cite{74}, quantum cryptography \cite{75}, quantum dense coding\cite{76}, quantum communication  \cite{77,78,79} , and quantum computation \cite{80,81}. Previously, it was believed that entanglement could explain all correlations. In order to measure the amount of entanglement, different criteria have been introduced \cite{82,83,84,85,86,87,88,89,90,91,92,93,94,95,96,97,98,99}. In spite of this, quantum entanglement does not cover everything about quantum correlations \cite{100,101}. Hence, it became imperative to establish an additional criterion to ensure the precise and comprehensive depiction of all quantum correlations. So far, quantum correlations have been quantified using a variety of measures, all of which have specific characteristics. Quantum discord QD is considered as one of the measurement criteria for quantum correlations \cite{102,103,104,105,106,107,108}.  In this work, quantum discord is used to show the quantum correlation between quantum memory $B$ and measured particle $A$. In a bipartite quantum system, the mutual information between subsystems A and B determines the total correlation
\begin{equation}\label{mu}
\mathcal{I}(\rho_{AB})=S(\rho_{A})+S(\rho_{B})-S(\rho_{AB}), 
\end{equation}
The state $\rho^{AB}$ is the hybrid object that contains both classical and quantum characteristic. According to Henderson and Vedral \cite{101}, correlation can also be divided into a quantum and a classical part. The classical part of correlation is defined as the maximum information that can be obtained about one subsystem through measurements on another subsystem. By considering the set of measurement operators $\lbrace P_k \rbrace$ that act on the subsystem $B$, the obtained information about the subsystem $A$ after the measurement resulting in outcome $k$ with probability $p_k$ can be defined as 
\begin{equation}\label{cla}
\mathcal{C}\left(\rho_{A B}\right)=\max _{\left\{P_k\right\}}\left[S\left(\rho_{A}\right)-\sum_k p_k S\left(\rho_{A \mid k}\right)\right].
\end{equation}
By subtracting classical correlation in Eq.\ref{cla} from quantum mutual information in Eq.\ref{mu}, quantum discord can be obtained as 
\begin{equation}
\mathcal{Q}(\rho_{A B})=\mathcal{I}(\rho_{AB})-\mathcal{C}\left(\rho_{A B}\right).
\end{equation}
There exist inverse relation between entropic uncertainty bound and quantum correlation between measured particle and quantum memory. In the case where maximal entangled state share between Alice and Bob, the entropic uncertainty bound is zero, and Bob can accurately predict Alice's measurement result. Hence, by maintaining the quantum correlation between Alice and Bob, Bob will be able to predict the result of Alice's measurement more accurately. In the real world, it is difficult to isolate quantum systems from their environment, and most quantum systems interact with their environment. So, the study of open quantum systems is of particular importance \cite{109}.  The evolution of open quantum systems is divided into two categories based on the direction of information flow during the evolution: Markovian and non-Markovian. In Markovian evolution information flow from system to environment continuously, while in non-Markovian evolution the back-flow of information will be appear during the evolution at some time interval.

As a result of the interaction of the quantum system with the environment, quantum properties such as quantum coherence or quantum correlation are changed and destroyed. In this work we will study the entropic uncertainty bound in the presence of quantum memory and quantum discord between measured particle and quantum memory when the quantum memory interacts with dissipative environment. The effect of the detuning between the transition frequency of a quantum memory and the center frequency of a cavity on entropic uncertainty bound and quantum discord between quantum memory and measured particle will be studied. In our study, both Markovian and non-Markovian regime will be considered.

\section{Dynamics of two-level open systems in the presence of detuning}
The system we're considering here consists of a two-level atom with transition frequency $\omega_0$ which, has coupled to an  environment includes of the quantized modes of high-Q cavity \cite{110,111}. The whole system can be described by following Hamiltonian
\begin{equation}
H=\frac{1}{2}\omega_0 \sigma_+ \sigma_- + \sum_k \omega_k b_k^{\dag} b_k+ \sum_k (g_k \sigma_+ b_k + g_k^{\star} \sigma_- b_k^{\dag}),
\end{equation}
where $\sigma_+$ and $\sigma_-$ are rising and lowering operators respectively. $\omega_k$ is the frequency of the $k$th field mode of cavity, $b_k$ and $b_k^{\dag}$ are annihilation and creation operator respectively. $g_k$ describes the coupling between atom and $k$th field mode of the cavity. If the total system consists of only one excitation, then the initial state can be expressed as 
\begin{equation}
\vert \psi(0) \rangle = c_1(0) \vert e \rangle_a \vert 0 \rangle_c + \sum_k c_k(0) \vert g \rangle_a \vert 1_k \rangle_c,
\end{equation}  
where $\vert e \rangle_a$ and $\vert g \rangle_a$ describe excited and ground states of the atom respectively, $\vert 0 \rangle_c$ represents the vacuum state of the cavity and $\vert 1_k \rangle_c$ is the state of the cavity with excitation at $k$th field mode of the cavity. So, the state of the whole system at time $t$ can be written as 
 \begin{equation}
\vert \psi(t) \rangle = c_1(t) \vert e \rangle_a \vert 0 \rangle_c + \sum_k c_k(t) \vert g \rangle_a \vert 1_k \rangle_c,
\end{equation}  
A series of differential equations for probability amplitudes can be derived by using the Schr$\ddot{o}$dinger equation as 
\begin{equation}
\begin{aligned}
& \dot{c}_1(t)=-i \sum_k g_k \exp \left[i\left(\omega_0-\omega_k\right) t\right] c_k(t), \\
& \dot{c}_k(t)=-i g_k^* \exp \left[-i\left(\omega_0-\omega_k\right) t\right] c_1(t) .
\end{aligned}
\end{equation}
Considering the initial assumption that the atom is in an excited state at the beginning i.e. $c_1(0)=1$ and $c_k(0)=0$, the following integro-differential equation can be obtane as 
\begin{equation}\label{int}
\dot{c}_1(t)=- \int_0^t dt_1 f(t-t_1)c_1(t).
\end{equation}
where $f(t-t_1$ is the correlation function that is associated to the spectral density of the cavity $J(\omega)$ by following relation
\begin{equation}
f(t-t_1)=\int d\omega J(\omega) \exp[i(\omega_0 - \omega)(t-t_1)].
\end{equation}
In the following the Lorentzian spectral density with detuning will be consider as 
\begin{equation}
J(\omega)=\frac{1}{2 \pi} \frac{\gamma \lambda^{2}}{(\omega_0-\omega-\delta)^{2}+\lambda^{2}},
\end{equation} 
where  $\delta=\omega_0 -\omega_c$ is the detuning between the transition frequency of the atom $\omega_0$ and the center frequency of the cavity $\omega_c$. As a result of detuning, the effective coupling between a qubit and the environment is decreased. For Lorentzian spectral density, the spectral width of the environment shows by $\lambda$ which associated with the correlation time of the environment as $\tau_c=\lambda^{-1}$. The time scale during which the state of the atom changes depends on parameter $\gamma$ through $\tau_a = \gamma^{-1}$ \cite{109}. By applying the Lorentzian spectral density, the correlation function of the environment $f(t-t_1)$ is obtained as follows
\begin{equation}
f(t-t_1)=\frac{1}{2}\gamma \lambda \exp[-(\lambda-i \delta)(t-t_1)].
\end{equation} 
Now, by substituting the above correlation function to Eq.\ref{int} and using Laplace transforms, the solution of  Eq.\ref{int} can be obtained as 
\begin{equation}
c_1(t)=G(t)c_1(0)
\end{equation}
where 
\begin{equation}
G(t)=e^{-(\lambda-i \delta) t / 2}\left[\cosh \left(\frac{\Omega t}{2}\right)+\frac{\lambda-i \delta}{\Omega} \sinh \left(\frac{\Omega t}{2}\right)\right],
\end{equation}
with $\Omega=\sqrt{(\lambda-i\delta)^2-2 \gamma \lambda}$. Additionally, the master equation that describes the dynamics of the model has the following form
\begin{equation}\label{master}
\frac{\partial}{\partial t} \rho(t)=\Gamma(t)\left(\sigma_-\rho(t) \sigma_{+}-\frac{1}{2}\left\{\sigma_{+} \sigma_{-}, \rho(t)\right\}\right).
\end{equation}
In above master equation the time-dependent decay rate $\Gamma(t)$ is given by
\begin{equation}
\Gamma(t)=\operatorname{Re}\left(\frac{2 \gamma \lambda \sinh (\Omega t / 2)}{\Omega \cosh (\Omega t / 2)+(\lambda-i \delta) \sinh (\Omega t / 2)}\right),
\end{equation}
where $Re$ means the real part of the value inside parentheses. By considering initial density matrix at time $t=0$, $\rho(0)=\sum_{i,j=1}^{2} \rho_{ij} \vert i \rangle \langle j \vert$, the solution of the master equation in Eq.\ref{master} can be obtain as 
 \begin{equation}
\rho(t)=\Phi_t \rho(0)=\left(\begin{array}{cc}
\rho_{11}|G(t)|^2 & \rho_{12} G(t) \\
\rho_{21} G^*(t) & 1-\rho_{11}|G(t)|^2
\end{array}\right)
\end{equation}
where $\Phi_t$ is quantum dynamical map which map initial state to the state at time $t$.  Using an operator-sum representation to represent the dynamics of the above two-level system is very instructive. The operator-sum representation of the quantum dynamical map $\Phi_t$ can be obtained as \cite{112}
\begin{equation}
\rho(t)=\Phi_t \rho(0)=\sum_{\alpha=1}^{2} K_\alpha(t) \rho(0) K_\alpha(t)^{\dag},
\end{equation} 
After straightforward calculations, related Kraus operators $K_\alpha(t)$ can be obtain as
\begin{equation}
K_1(t)=\left(\begin{array}{cc}
G(t) & 0 \\
0 & 1
\end{array}\right), \quad K_2(t)=\left(\begin{array}{cc}
0 & 0 \\
\sqrt{1-|G(t)|^2} & 0
\end{array}\right).
\end{equation}
\section{Quantum correlations and entropic uncertainty bound with detuning}
\begin{figure}[H]
    \centering
  \includegraphics[width = 0.75\linewidth]{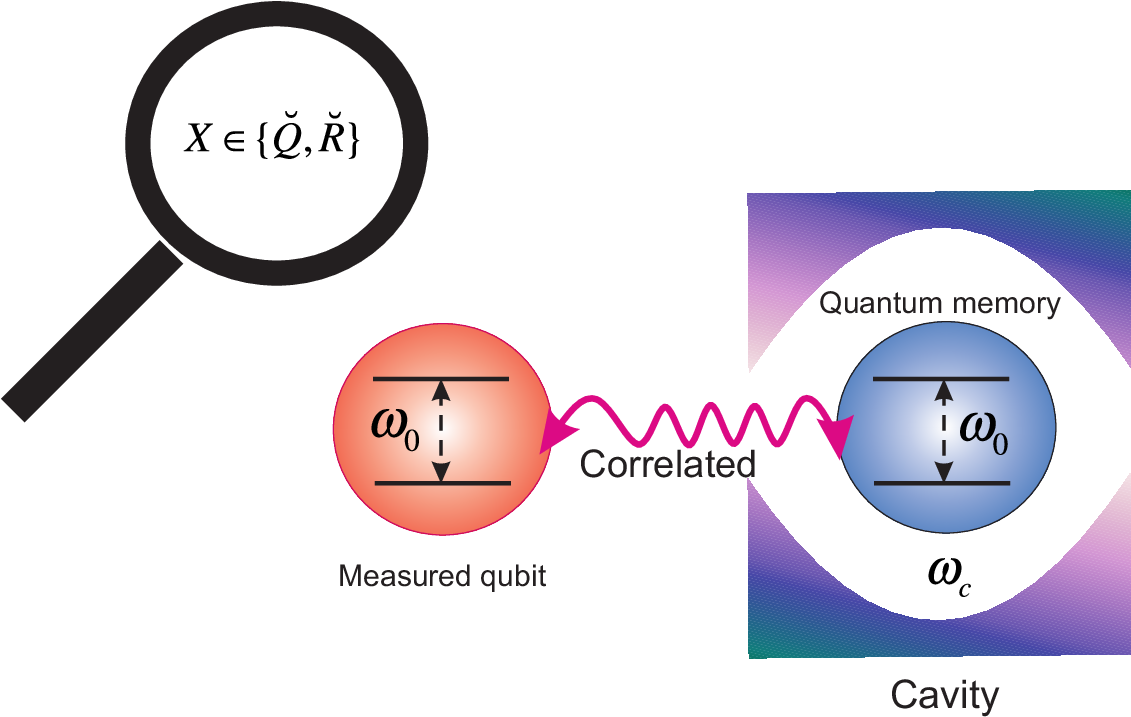}
    \centering
    \caption{Schematic representation of the model, where the quantum memory interacts with dissipative environment, while measured qubit is free and Alice does her measurement $X \in \lbrace Q, R \rbrace$ on it. 
}\label{Fig1}
\end{figure}

In this section, the effects of detuning on quantum correlations and entropic uncertainty bound will be studied. Here, we are considering the scenario in which the quantum memory $B$ interacts with the environment while the measured particle $A$ remains unaffected by environment and Alice does her measurement on it. The dynamics of bipartite state $\rho^{AB}$ can be described as
\begin{equation}
\rho^{AB}(t)=\sum_{\alpha=1}^{2} (\mathrm{I}\otimes K_{\alpha}(t))\rho^{AB}(0)(\mathrm{I}\otimes K_{\alpha}(t))^{\dag}.
\end{equation}
The schematic of the scenario has been sketched in Fig.\ref{Fig1}.    

Let us consider the case in which the following two-qubit state has been shared between Alice and Bob as 
\begin{equation}\label{state}
\rho_\psi(r, \theta)=r|\psi(\theta) \rangle \langle \psi(\theta)|+\frac{1-r}{4} \mathrm{I} \otimes \mathrm{I}
\end{equation}
where $r$ quantify the purity of the initial state and 
\begin{equation}
|\psi(\theta)\rangle=\cos \theta|00\rangle+\sin \theta|11\rangle,
\end{equation}
parameter $\theta$ is considered the correlation parameter. In Eq.\ref{state} first qubit is considered as measured particle and the second one is considered a quantum memory. QD has its maximum value one for $\theta=\pi/4$ and $3 \pi/4$ with $r=1$. 

 Let us consider the case in which the quantum memory is affected by cavity. By considering above initial state the evolved density matrix can be obtained as
\begin{equation}\label{DMS}
\rho _{\text{AB}}(t)=\left(
\begin{array}{cccc}
 \rho _{11} & 0 & 0 & \rho _{14} \\
 0 & \rho _{22} & 0 & 0 \\
 0 & 0 & \rho _{33} & 0 \\
 \rho _{41} & 0 & 0 & \rho _{44} \\
\end{array}
\right)
\end{equation}
where 
\begin{eqnarray}
\rho_{11}&=&\frac{1}{4} \vert G(t) \vert^{2} (1+r+2r \cos 2 \theta), \nonumber \\ 
\rho_{22}&=&\frac{1}{4}(1-r+ \vert 1-G(t)^2 \vert (1+r+2r \cos 2 \theta)), \nonumber \\
\rho_{33}&=&\frac{1}{4}\vert G(t) \vert ^2 (1-r), \nonumber \\
\rho_{44}&=&\frac{1}{4}(1+r+(1-r)\vert 1-G(t)^2 \vert -2 r \cos 2 \theta), \nonumber \\
\rho_{14}&=&\rho_{41}^{\star}=G(t)r \cos \theta \sin \theta, 
\end{eqnarray}
\subsection{Entanglement and Quantum discord}
Several criteria are used in the measurement of quantum correlations, as mentioned in the introduction. In this study, we measure quantum correlations using two practical and optimal criteria: concurrence(entanglement measure) and QD. The concurrence for two-qubit states can be defined as \cite{113}
\begin{equation}
\mathcal{C}(\rho_{AB})=\max \lbrace0, \varepsilon_1-\varepsilon_2-\varepsilon_3-\varepsilon_4 \rbrace,
\end{equation} 
where $\varepsilon_i$'s are the eigenvalues of  $R=\sqrt{\sqrt{\rho}\tilde{\rho}\sqrt{\rho}}$, in decreasing order and $\tilde{\rho}=(\sigma_y \otimes \sigma_y)\rho^{\star}(\sigma_y \otimes \sigma_y)$ with $\sigma_y$ is the $y$ component of Pauli matrices and $\rho^{\star}$ is the complex conjugate of $\rho$. The concurrence of the state in Eq.\ref{DMS} can be obtained as 
\begin{equation}
\mathcal{C}(\rho)=2 \max \lbrace0, \vert \rho_{14} \vert -\sqrt{\rho_{22}\rho_{33}}\rbrace.
\end{equation}
In addition to concurrence, QD can also be obtained in the following way \cite{103,104}
\begin{equation}
\mathcal{Q}(\rho_{AB})=\min \lbrace \mathcal{Q}_1, \mathcal{Q}_2\rbrace
\end{equation}
with 
\begin{equation}
\mathcal{Q}_j=h(\rho_{11}+\rho_{33})+\sum_{i=1}^{4} \lambda_i \log_2 \lambda_i  + \mathcal{D}_j
\end{equation}
where $j \in \lbrace 1,2\rbrace$ and
\begin{equation}
\begin{aligned}
& \mathcal{D}_1=h\left(\frac{1+\kappa}{2}\right), \\
& \mathcal{D}_2=-\sum_i \rho_{i i} \log _2 \rho_{i i}-h\left(\rho_{11}+\rho_{33}\right),
\end{aligned}
\end{equation}
in above equations we have $\kappa=\sqrt{[1-2(\rho_{33}+\rho_{44})]^2 + 4 \vert \rho_{14} \vert^{2}}$ and $h(x)=-x \log _2 x-(1-x) \log _2(1-x)$.
\subsection{Entropic uncertainty bound}
Here, two incompatible observables are chosen $Q=\sigma_x$ and $R=\sigma_z$ where $\sigma_x$ and $\sigma_z$ are $x$ and $z$ component of Pauli operators. By choosing these observables the complementarity parameter is obtained as $c=1/2$.  According to the selection of these observables, the Holevo quantities corresponding to each of them are obtained as follows 
\begin{equation}
I\left(\sigma_z ; B\right)=h(\rho_{11}+\rho_{22})+h(\rho_{11}+\rho_{33})+\sum_i \rho_{i i} \log _2 \rho_{i i},
\end{equation}
and 
\begin{equation}
I\left(\sigma_x ; B\right)=1+h(\rho_{11}+\rho_{33})+\sum_i \xi_i \log _2 \xi_i,
\end{equation}
where $\xi_1=\xi_2=\frac{1-k}{4}$ and $\xi_3=\xi_4=\frac{1+k}{4}$ with $k=\sqrt{4\left|\rho_{14} \right|^2+\left[1-2\left(\rho_{22}+\rho_{44}\right)\right]^2}$. From Eq. \ref{DMS}, the mutual information $\mathcal{I}(\rho_{AB}(t))$ can be obtained as 
\begin{equation}
I(A;B)=h(\rho_{11}+\rho_{22})+h(\rho_{11}+\rho_{33})+\sum_i \lambda_i \log _2 \lambda_i,
\end{equation}
where $\lambda_i$'s are the eigenvalues of $\rho_{AB}(t)$. So, EUB (right hand side of Eq.\ref{Adabi}) associated with measurement of two observables $\sigma_x$ and $\sigma_z$, can be obtained as 
\begin{equation}
EUB \equiv 1+S(A \mid B)+\max \left\{0, I(A ; B)-I\left(\sigma_x ; B\right)-I\left(\sigma_z ; B\right)\right\},
\end{equation}
\section{Results}
\begin{figure}[H]
    \centering
  \includegraphics[width = 0.75\linewidth]{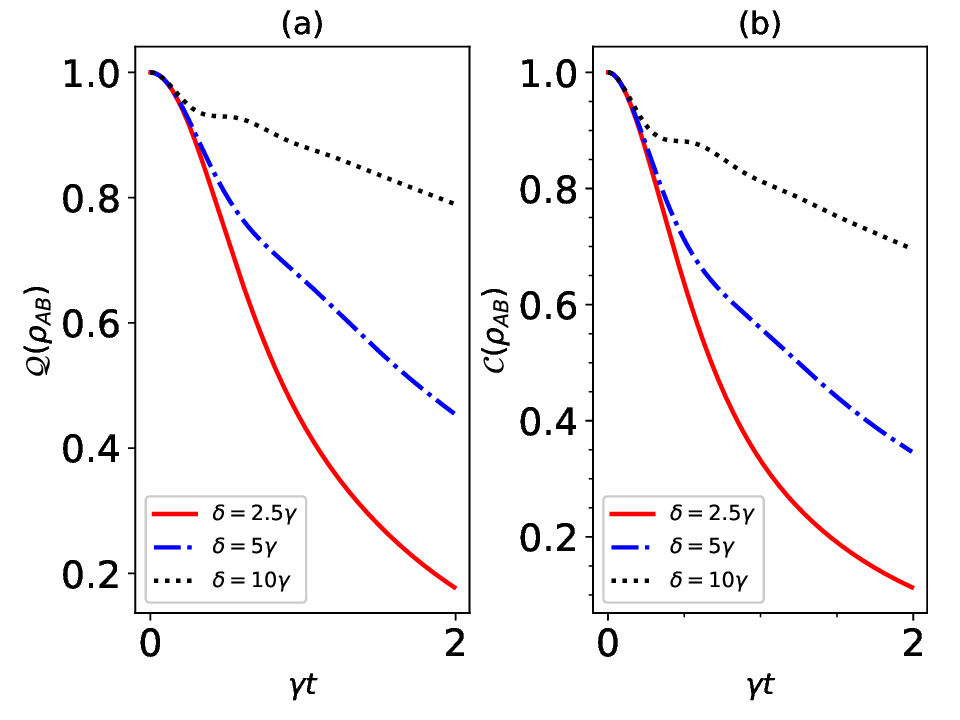}
    \centering
    \caption{(a)Quantum discord as a funcition of dimensionless parameter $\gamma t$ in Markovian regime with $\lambda =10 \gamma$ for different values of detuning $\delta$, for maximally entangled initial state $r=1,\theta=\pi/4$. (b) concurrence as a funcition of dimensionless parameter $\gamma t$ in Markovian regime with $\lambda =10 \gamma$ for different values of detuning $\delta$, for maximally entangled initial state $r=1,\theta=\pi/4$.
}\label{Fig2}
\end{figure}

\begin{figure}[H]
    \centering
  \includegraphics[width = 0.75\linewidth]{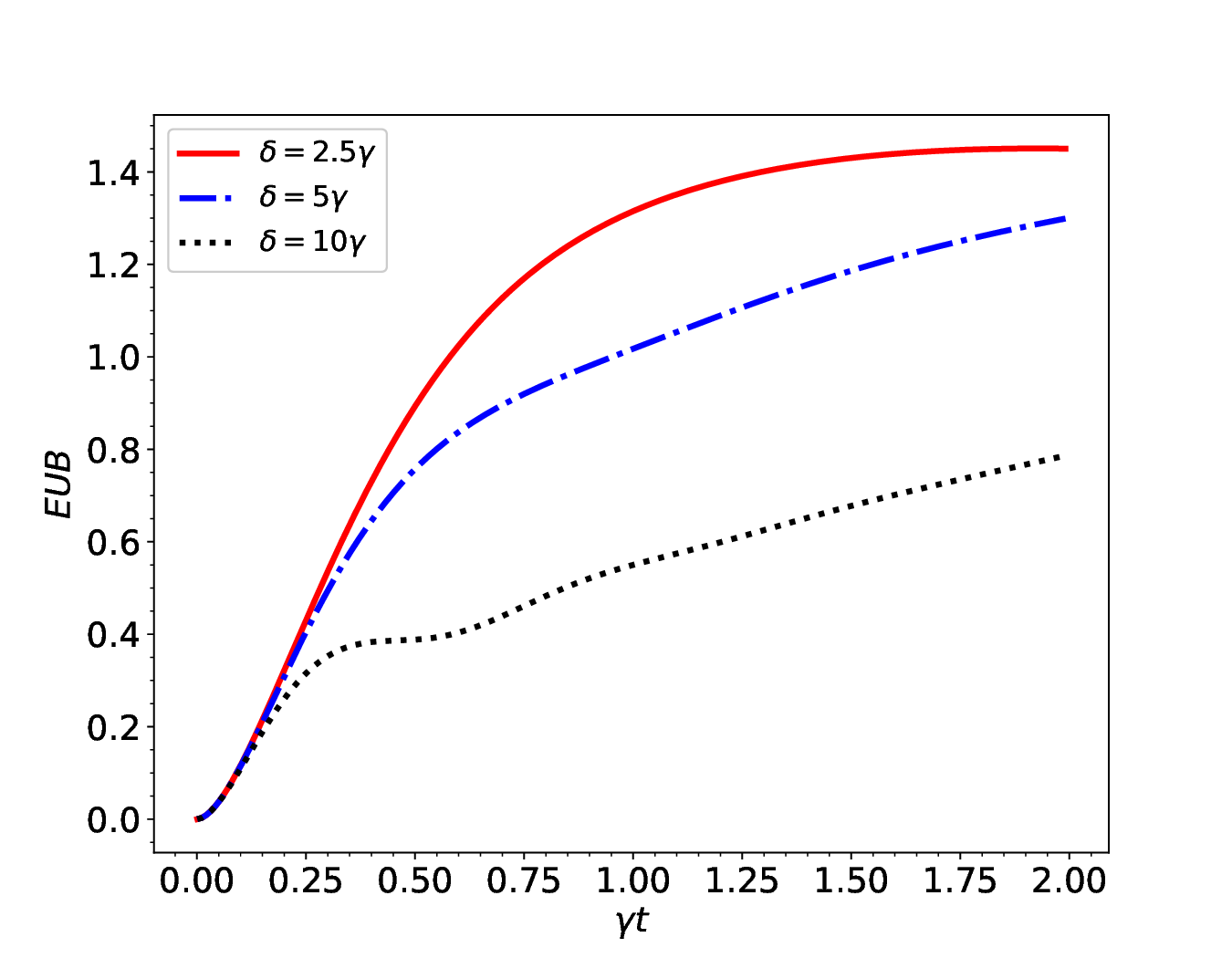}
    \centering
    \caption{Entropic uncertainty bound as a funcition of dimensionless parameter $\gamma t$ in Markovian regime with $\lambda =10 \gamma$ for different values of detuning $\delta$, for maximally entangled initial state $r=1,\theta=\pi/4$ 
}\label{Fig3}
\end{figure}

\begin{figure}[H]
    \centering
  \includegraphics[width = 0.75\linewidth]{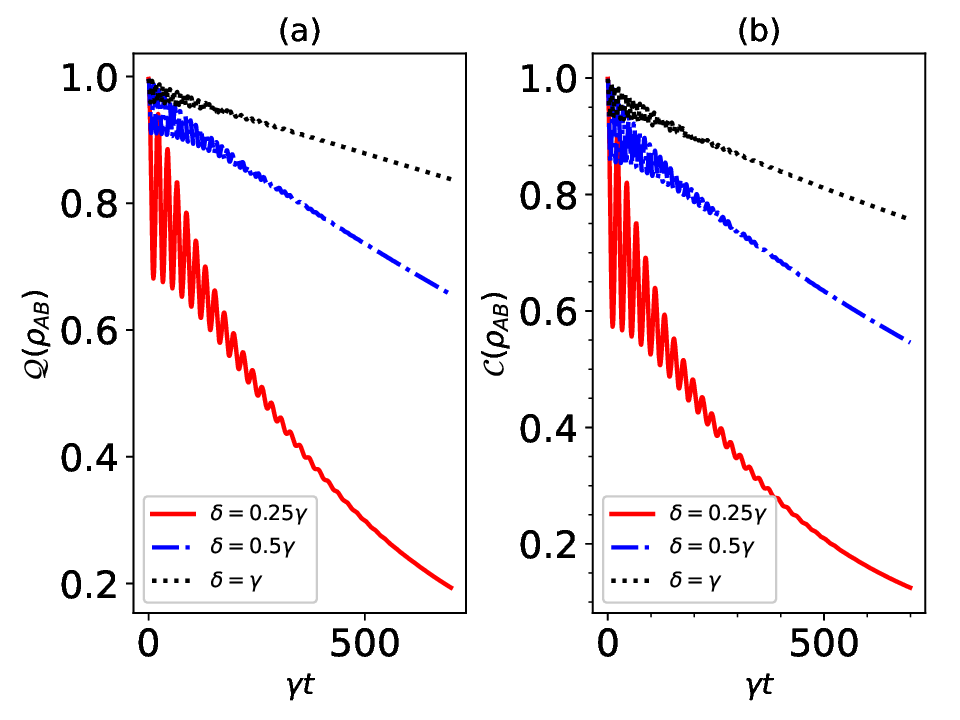}
    \centering
    \caption{(a)Quantum discord as a funcition of dimensionless parameter $\gamma t$ in non-Markovian regime with $\lambda =0.1 \gamma$ for different values of detuning $\delta$, for maximally entangled initial state $r=1,\theta=\pi/4$. (b) concurrence as a funcition of dimensionless parameter $\gamma t$ in non-Markovian regime with $\lambda =0.1 \gamma$ for different values of detuning $\delta$, for maximally entangled initial state $r=1,\theta=\pi/4$. 
}\label{Fig4}
\end{figure}

\begin{figure}[H]
    \centering
  \includegraphics[width = 0.75\linewidth]{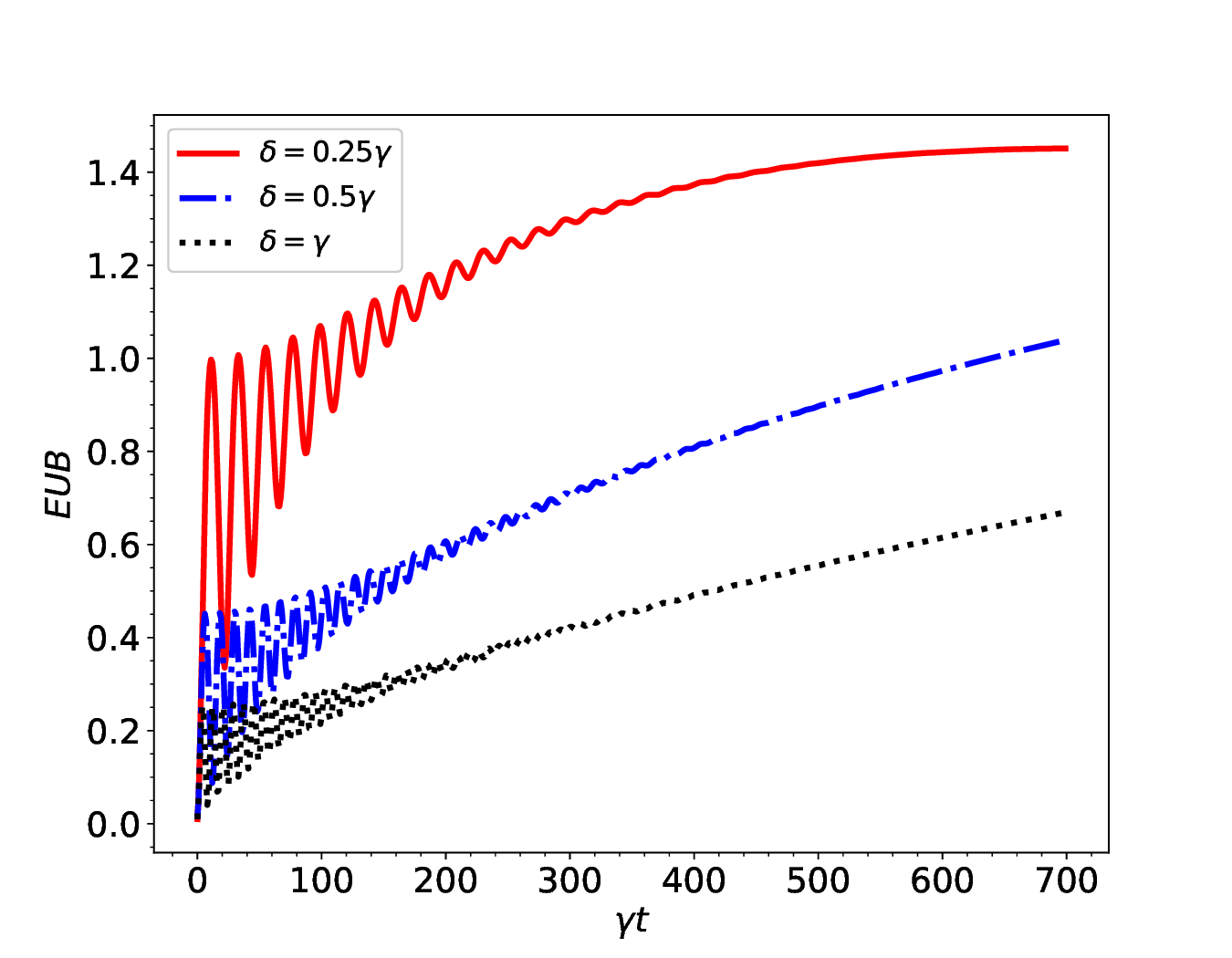}
    \centering
    \caption{Entropic uncertainty bound as a funcition of dimensionless parameter $\gamma t$ in non-Markovian regime with $\lambda =0.1 \gamma$ for different values of detuning $\delta$, for maximally entangled initial state $r=1,\theta=\pi/4$. 
}\label{Fig5}
\end{figure}

In this section we study the behavior of the EUB and quantum correlation in the presence of detuning in dissipative environment. We consider the maximally entangled state with $r=1$, $\theta=\pi/4$ as the initial state which has been shared between ALice and Bob. 
By setting $r=1$ and $\theta=\pi /4$  in Eq. \ref{state}, we have the maximally entangled initial state which has been shared between Alice and Bob. In Fig.\ref{Fig2}, quantum correlations has been plotted as a function of dimensionless parameter $\gamma t$ in Markovian regime with $\lambda=10 \gamma$ for different values of detuning $\delta$. Fig.\ref{Fig2}.(a), shows the QD in Markovian regime for different values of detuning $\delta$. It can be seen that increasing $\delta$ can protect the QD from decay. In other words, increasing detuning parameter  leads to more preservation of QD in the presence of the dissipative environment. In Fig.\ref{Fig2}(b), concurrence has been plotted as a function of dimensionless parameter $\gamma t$. As can be seen the same results have been obtained for concurrence. So, it can be concluded that increasing detuning can protect quantum correlation from decoherence.   
As mentioned before, there exist inverse relation between EUB and quantum correlation between measured particle and quantum memory. So, it is expected increasing detuning parameter leads to decreasing in EUB. In Fig.\ref{Fig3}, the EUB has been plotted as a function of dimensionless parameter $\gamma t$ in Markovian regime with $\lambda=10 \gamma$. As can be seen, the EUB decreases with increasing the detuning parameter. This result is in agreement with the behavior of quantum correlations in dissipative environment with detuning.    
In Fig.\ref{Fig4}, entanglement and QD have been plotted as a function of dimensionless parameter $\gamma t$ for different values of detuning parameter in non-Markovian regime with $\lambda=0.1 \gamma$ for maximally entangled initial state. Fig.\ref{Fig4}(a), shows QD in terms of dimensionless for different values of detuning parameter in non-Markovian regime. As can be seen, QD can be protected from decoherence during the evolution by increasing detuning parameter. In Fig.\ref{Fig4}(b), the concurrence as a measure for entanglement has been plotted as a function of detuning parameter in non-Markovian regime for different values of detuning parameter. Again, one can see the concurrence can be preserve during interaction with dissipative environment by increasing the detuning parameter.  
In Fig.\ref{Fig5}, the EUB has been plotted as a function of dimensionless parameter $\gamma t$ in non-Markovian environment for different values of detuning parameters $\delta$. As expected, it can be seen that the EUB decreases with increasing the detuning parameter. 

\section{Conclusion}
In this work, the effects of detuning on quantum correlations and entropic uncertainty bound have been studied. It was observed that there exist inverse relation between entropic uncertainty bound and quantum correlation between measured particle and quantum memory. It was also shown that the quantum correlation increases with increasing detuning  between the transition frequency of a quantum memory and the center frequency of a cavity. The situation for entropic uncertainty bound was quite different. It was observed that the entropic uncertainty bound can be reduced with increasing the detuning parameter. From these results, it can be concluded that the correlation between the measured particle, which is in the possession of Alice and quantum memory , which is in the possession of Bob, has an important role in guessing the result of Alice's measurement by Bob. 
\section*{Data availability}
No datasets were generated or analyzed during the current study.


\section*{Competing interests}
The authors declare no competing interests.

\section*{ORCID iDs}
Maryam Hadipour \\
https://orcid.org/0000-0002-6573-9960 \\
Soroush Haseli\\
https://orcid.org/0000-0003-1031-4815

\end{document}